\documentclass[preprint,12pt,hyphens]{elsarticle}

\usepackage{amssymb}

\usepackage{amsmath}
\usepackage{color,array}
\usepackage{graphicx}
\usepackage{booktabs}
\usepackage{amsmath}
\usepackage{multirow}
\usepackage{subcaption}
\usepackage[numbers]{natbib}

\usepackage{enumitem}
\usepackage{tcolorbox}
\usepackage[colorlinks=true, citecolor=blue]{hyperref}
\usepackage[a4paper, total={7in, 8in}]{geometry}

\usepackage{colortbl}
\usepackage{lscape}
\usepackage{setspace}
\usepackage{graphicx}
\usepackage{graphics}
\usepackage{wrapfig}
\usepackage{amsmath}
\usepackage{tabularx}
\usepackage[table,dvipsnames]{xcolor}
\definecolor{arsenic}{rgb}{0.23, 0.27, 0.29}
\usepackage[font=small,skip=5pt]{caption}
\usepackage{titlesec}
\usepackage{arydshln}
\usepackage{soul}
\usepackage{subcaption}
\usepackage{adjustbox}
\usepackage{lineno}
\usepackage{setspace}
\usepackage{multirow}
\usepackage{multicol}
\usepackage{array}
\usepackage{makecell}
\usepackage{MnSymbol}
\usepackage{mathtools}
\usepackage{scrextend}
\usepackage{caption}
\usepackage{enumitem}   
\definecolor{lightyellow}{RGB}{255, 255, 204}
\usepackage{siunitx}
\tcbset{
  colback=gray!10,    
  colframe=gray!50,   
  boxrule=0.5pt,
  arc=2mm,            
  fonttitle=\bfseries,
  coltitle=black
}
\journal{the journal}

\begin{document}

\begin{frontmatter}

\title{Mapping the Urban Mobility Intelligence Frontier: A Scientometric Analysis of Data-Driven Pedestrian Trajectory Prediction and Simulation}

\author[inst1]{Junhao Xu\corref{corauth}} 
\author[inst1]{Hui Zeng} 
\affiliation[inst1]{organization={School of Engineering, Guangzhou College of Technology and Business},
            city={Guangzhou},
            country={China}}
\cortext[corauth]{Corresponding author. Email: xujunhao2@gzgs.edu.cn; xujunhao1105@gmail.com}
\begin{abstract}
Understanding and predicting pedestrian dynamics has become essential for shaping safer, more responsive, and human-centered urban environments. This study conducts a comprehensive scientometric analysis of research on data-driven pedestrian trajectory prediction and crowd simulation, mapping its intellectual evolution and interdisciplinary structure. Using bibliometric data from the Web of Science Core Collection, we employ SciExplorer and Bibliometrix to identify major trends, influential contributors, and emerging frontiers. Results reveal a strong convergence between artificial intelligence, urban informatics, and crowd behavior modeling—driven by graph neural networks, transformers, and generative models. Beyond technical advances, the field increasingly informs urban mobility design, public safety planning, and digital twin development for smart cities. However, challenges remain in ensuring interpretability, inclusivity, and cross-domain transferability. By connecting methodological trajectories with urban applications, this work highlights how data-driven approaches can enrich urban governance and pave the way for adaptive, socially responsible mobility intelligence in future cities.
\end{abstract}



\end{frontmatter}

\section{Introduction}
Pedestrian trajectory prediction and simulation is an interdisciplinary research field where data-driven models, particularly machine learning and deep learning techniques, are employed to model, predict, and simulate human movement dynamics in diverse environments \cite{1,2,3}. Although research on pedestrian dynamics can be traced back to the seminal social force model \cite{4}, the advent of large-scale mobility datasets and sensing technologies has substantially transformed the landscape in recent decades. Modern data sources, ranging from video surveillance \cite{5}, wearable sensors \cite{6}, to mobile and GPS data \cite{7}, have enabled the development of more accurate and context-aware predictive models.

The societal importance of pedestrian trajectory modeling is increasingly evident in critical applications such as traffic simulation \cite{8}, smart city planning \cite{9}, and public safety management \cite{10}. For instance, urban traffic simulations now incorporate fine-grained pedestrian movement patterns to optimize intersections, manage crowd flows in transit hubs, and enhance evacuation planning in emergency scenarios \cite{11}. In smart cities, understanding pedestrian mobility is essential for designing safer, more efficient urban spaces, while in public safety contexts, predicting crowd behavior can help mitigate risks in mass gatherings or critical incidents \cite{12}.

Despite its longstanding foundations, the field is dynamic and continuously reshaped by emerging computational techniques and data modalities. The incorporation of graph neural networks (GNNs) \cite{13} and transformer-based models \cite{14} represents methodological advancements that push the boundaries of predictive accuracy and interpretability in trajectory forecasting. Additionally, the growing interest in multimodal approaches—combining vision, location, and contextual data—reflects an ongoing shift towards more holistic modeling paradigms \cite{15}.

Given the rapid evolution of methods and the multidisciplinary nature of pedestrian trajectory research, it is challenging to delineate the precise contours of this field. A quantitative and systematic understanding of its development can inform researchers, urban planners, policy-makers, and other stakeholders about the prevailing trends and emerging frontiers. Scientometric analysis, leveraging bibliometric data from platforms like Web of Science, offers a robust means to map the intellectual structure, collaboration networks, and thematic evolution of this research domain \cite{16}. 

Although pedestrian trajectory prediction and simulation have been extensively studied, there is still a lack of comprehensive scientometric analysis that systematically maps the development of this data-driven field. To address this gap, we conduct a quantitative analysis of publications indexed in the Web of Science up to 2025, examining key research trends, international collaboration networks, and prominent thematic areas. These developments extend beyond technical modeling—they form part of the growing movement toward urban intelligence, where data-driven systems interact with human dynamics to improve city resilience, inclusivity, and sustainability.

\section{Methods}
\subsection{Study Aims}
The design of a bibliometric study should be tailored to its aims \cite{17}. In this study, our aim is to answer the following core research questions on interdisciplinarity,internationality, collaboration, and trending topics, respectively:

\begin{enumerate}[label=R\arabic*]
    \item Is data-driven pedestrian trajectory prediction and simulation an emerging and promising research direction?
    \item Which academic disciplines have influenced the development of data-driven pedestrian trajectory prediction and simulation, and conversely, which disciplines have been influenced by it? In other words, who is this field serving, and how is it positioned within the broader scientific landscape?
    \item What are the most influential publication venues (journals and conferences), countries, and leading authors in the field of data-driven pedestrian trajectory prediction and simulation, and how have their contributions shaped the development and global evolution of this research domain?
    \item What are the major research hotspots in data-driven pedestrian trajectory prediction and simulation. How have key papers shaped the intellectual trajectory and paradigm shifts within this field?

    \item  What are the emerging frontier technologies or methodological paradigms under development that have the potential to become the next milestone in data-driven pedestrian trajectory prediction and simulation?
\end{enumerate}

Scientometric analysis serves as a lens through which the evolution of urban mobility intelligence research can be quantitatively traced and contextualized within the broader ecosystem of urban informatics.
\subsection{Journal selection}
We searched the Web of Science Core Collection on 2025-08-24 (inclusive) using the following topic search (TS) string in the 'Topic' field:
Indexes: Web of Science Core Collection (All indexes). Retrieved on 2025-08-24.
Exported fields: Full record + cited references (including Authors, Title, Source, Year, Abstract, Keywords, Author Affiliations, Times Cited).
\begin{tcolorbox}
TS=(("pedestrian trajectory" OR "pedestrian movement" OR "human trajectory" OR "crowd trajectory" OR "crowd movement" OR "pedestrian path prediction" OR "pedestrian flow simulation" OR "pedestrian behavior simulation" OR "crowd simulation" OR "crowd dynamics simulation")
  AND
  ("data-driven" OR "machine learning" OR "deep learning" OR "artificial intelligence" OR "graph neural network" OR "reinforcement learning" OR "supervised learning" OR "unsupervised learning" OR "neural network" OR "generative model")).
\end{tcolorbox}
We retrieved a total of 572 publications, including both journal and conference papers. Among these, 12 were review articles, which were excluded from further analysis to avoid duplication of synthesized knowledge. Consequently, 560 original research papers remained and were retained for subsequent screening and detailed analysis. The filtering process focused on identifying studies that directly addressed pedestrian and crowd trajectory modeling, prediction, or simulation using data-driven and machine learning approaches. These selected papers formed the core dataset for our bibliometric and content analysis in the next stage.

\subsection{Data Analysis Tools}
For the analysis of the 560 selected papers, we employed a combination of bibliometric and text-mining tools to extract and visualize research trends. Specifically, we utilized \textbf{SciExplorer}\footnote{\url{https://smartdata.las.ac.cn/SciExplorer/?lang=CN}} and the \textbf{Bibliometrix}\footnote{\url{https://github.com/massimoaria/bibliometrix}} package in R to conduct comprehensive bibliometric analysis. These tools were jointly applied to process and visualize bibliographic data exported from the Web of Science, enabling the exploration of publication patterns, keyword co-occurrence networks, and thematic evolution across years. By integrating the interactive visualization features of SciExplorer with the statistical analysis capabilities of Bibliometrix, we were able to ensure both interpretability and analytical depth in identifying research hotspots and intellectual structures within the domain of data-driven pedestrian and crowd trajectory studies.

\section{Results}
This section presents detailed visualizations and analytical results corresponding to each research question.

\subsection{R1}

\begin{figure}
    \centering
    \includegraphics[width=\linewidth]{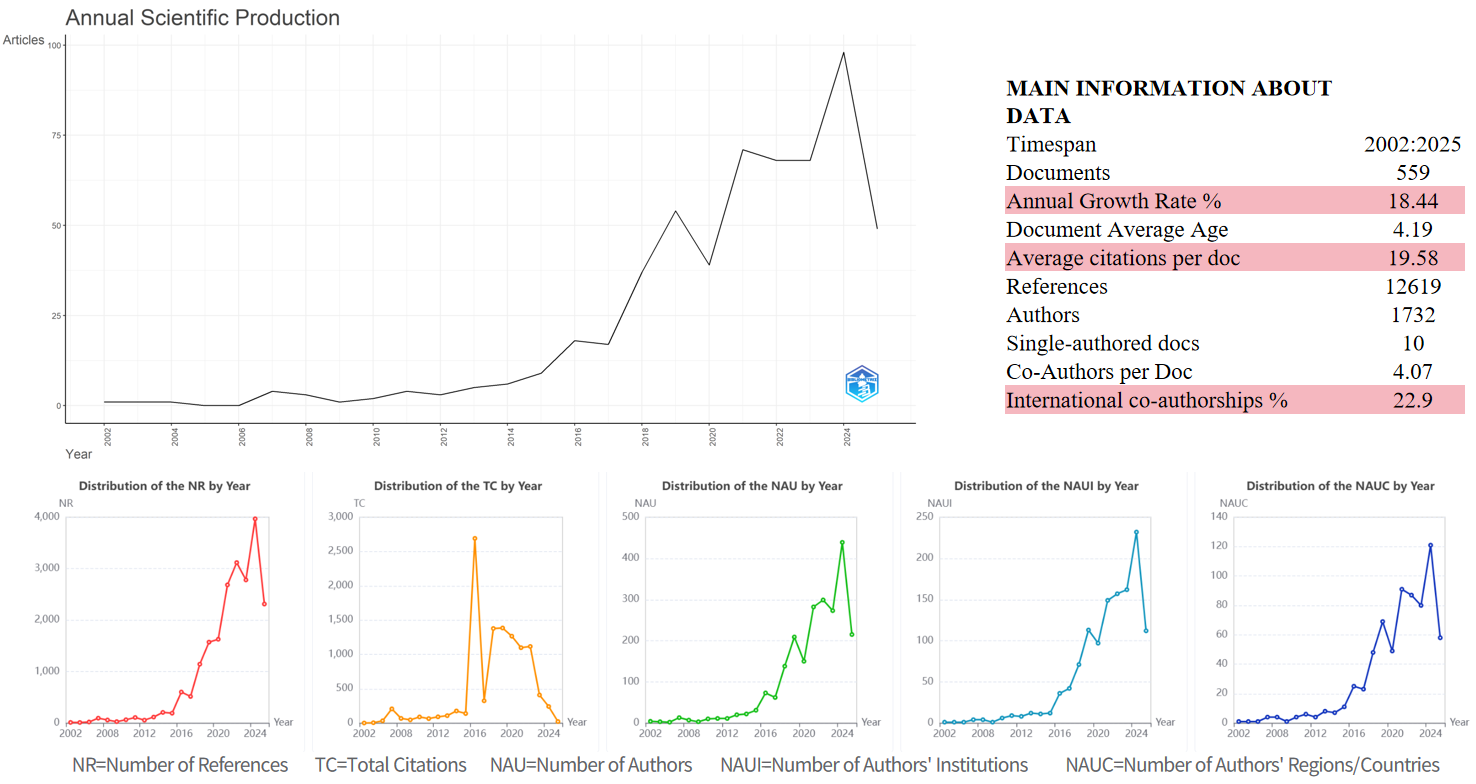}
    \caption{Literature statistics overview.}
    \label{fig:1}
\end{figure}

The evidence presented in Figure \ref{fig:1} strongly supports the assertion that data-driven pedestrian trajectory prediction and simulation is indeed an emerging and promising research direction. The annual growth rate of 18.44\%, coupled with the noticeable surge in the number of publications since 2017, clearly indicates a growing interest in the field. This is further corroborated by the sharp increase in references and citations, demonstrating not only an increase in research output but also an escalating academic impact.

The sharp upward trend in annual publications, coupled with a relatively young document average age of 4.19 years, highlights a field that is gaining momentum in recent years. This rapid expansion is a hallmark of a burgeoning research area where researchers are increasingly exploring new techniques, algorithms, and applications. The influx of articles since 2017 aligns with the integration of advanced data-driven techniques such as machine learning and deep learning, which have reshaped how pedestrian movement and crowd behavior are modeled and predicted. The average citations per document of 19.58 and the total citations distribution indicate that the research outputs in this field are not only increasing in quantity but are also gaining significant academic recognition. High citation counts are a strong indicator of the research's relevance and contribution to advancing the understanding of complex systems like pedestrian movement.  One of the most striking aspects of this field is its collaborative nature, as evidenced by the average co-authors per document of 4.07. This suggests that pedestrian trajectory prediction and simulation is a highly interdisciplinary research area. The increase in the number of authors, authors' institutions, and regions/countries participating in this research highlights the global and interdisciplinary collaboration that is fundamental to its growth. The 22.9\% international co-authorship rate further emphasizes that this field is not isolated but is part of a global research network, where international contributions and cross-border collaboration are playing a significant role in its development. This level of collaboration is essential for addressing complex, multi-faceted challenges, such as simulating pedestrian behavior in diverse environmental contexts, optimizing urban infrastructure, and enhancing public safety in smart cities. 

The substantial growth in publications, citations, and collaborative efforts—along with its increasing impact on urban planning and public safety strongly positions data-driven pedestrian trajectory prediction and simulation as a promising research direction. The field is not only emerging rapidly but is also expanding its influence across various disciplines, making it a central area for future research. As data sources, computational methods, and interdisciplinary collaboration continue to evolve, the potential for this field to shape urban mobility and smart city solutions becomes increasingly evident. This signifies a mature and highly relevant area of research with broad societal and academic applications.

\subsection{R2}
To address this research question, we employed a science mapping approach to analyze the disciplinary structure of the field. Specifically, we retrieved all the filtered publications from the database and categorized them into 12 distinct academic disciplines. We then performed a statistical analysis to quantify the disciplinary distribution and interconnections among these publications.

In the resulting visualization, each color represents a specific academic discipline, allowing for a clear distinction between fields. Each circle within the map corresponds to a thematic cluster under its respective discipline, with the size of the circle indicating the volume of publications associated with that theme — larger circles represent a higher number of publications. The lines connecting the circles represent co-occurrence relationships between themes, highlighting how often different topics are studied together within the same publications. The thickness or density of these lines reflects the strength of association between themes. This visual representation provides an intuitive understanding of the interdisciplinary landscape of data-driven pedestrian trajectory prediction and simulation. It not only reveals the dominant disciplines contributing to this field but also uncovers how various topics interrelate and form clusters of research activity.

\begin{figure}[!htbp]
    \centering
    \begin{subfigure}[b]{0.85\linewidth}
        \centering
        \includegraphics[trim=0cm 3cm 0cm 2.5cm, clip, width=\linewidth]{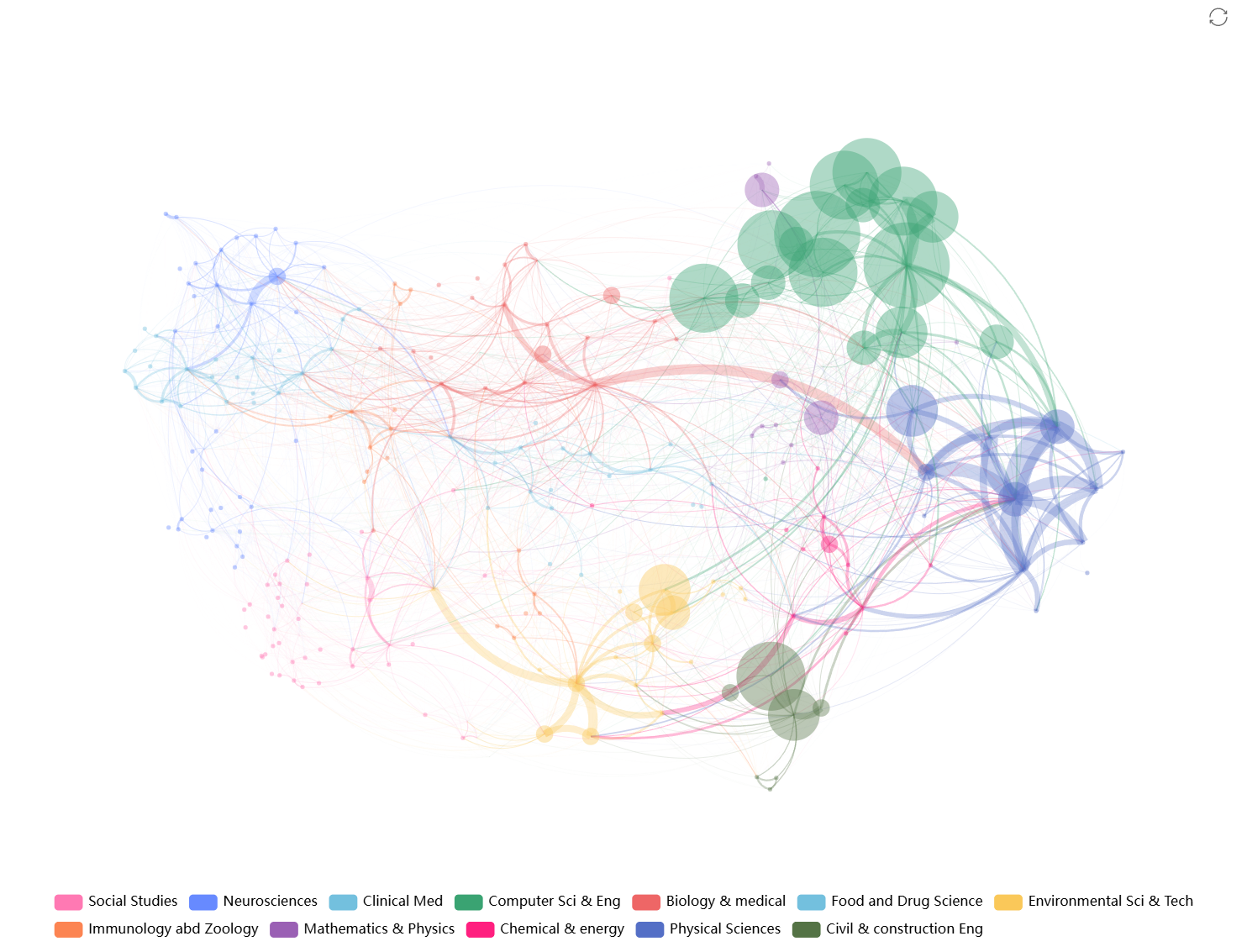}
        \label{fig:subfig-a}
    \end{subfigure}
    

    \begin{subfigure}[b]{0.85\linewidth}
        \centering
        \includegraphics[trim=0cm 0cm 0cm 2.5cm, clip, width=\linewidth]{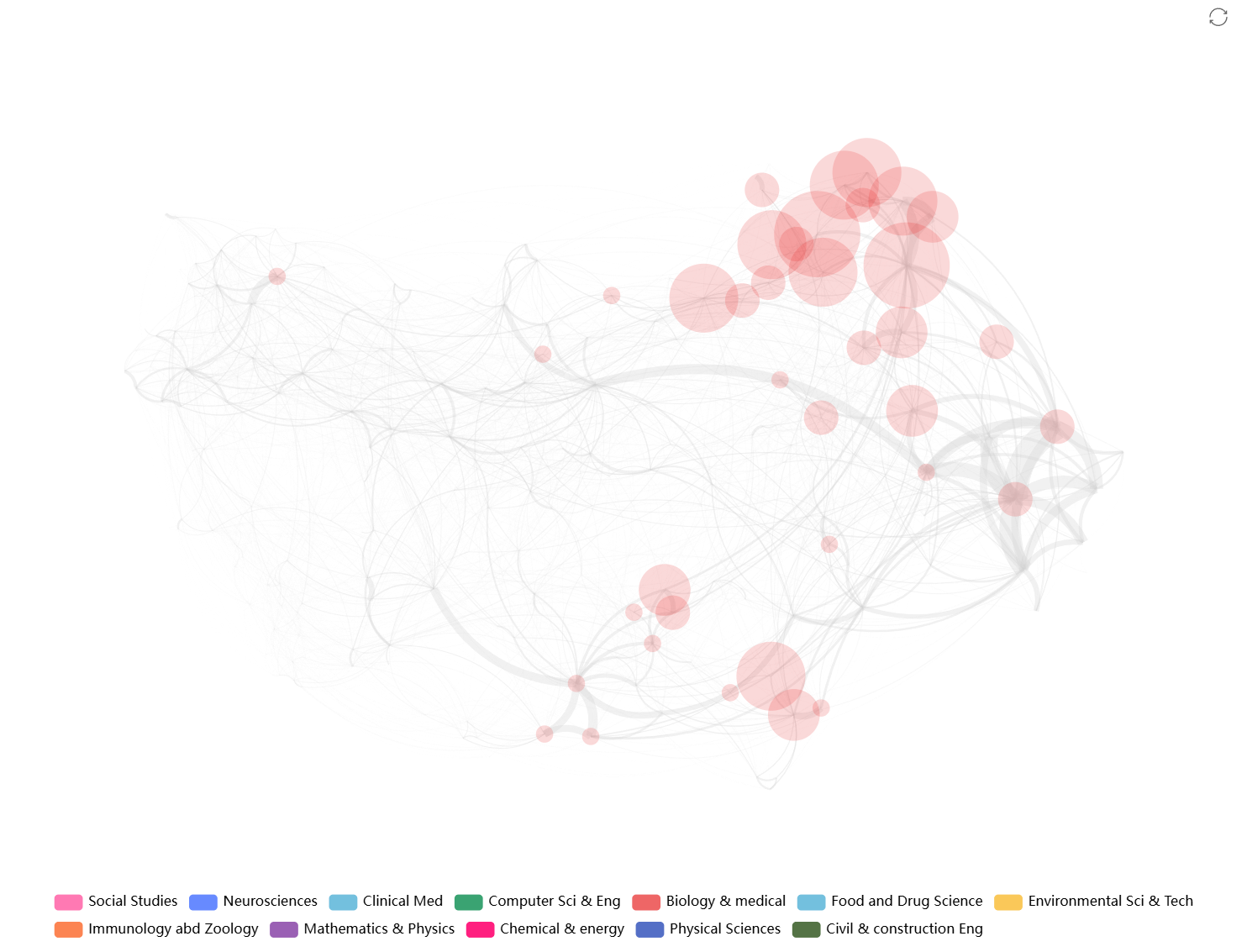}
        \label{fig:subfig-b}
    \end{subfigure}

    \caption{Global Map of Science, with different colors representing distinct disciplines.}
    \label{fig:enter-label}
\end{figure}

The upper part of Figure \ref{fig:enter-label} presents the complete disciplinary clustering map, while the lower part shows a density map of disciplines generated by applying a threshold filter. From the visualization, we can observe that Computer Science and Engineering, Civil and Construction Engineering, Mathematics \& Physics, Physical Sciences, and Environmental Science and Technology form prominent clusters in the upper part of the map. This indicates that these disciplines have contributed significantly to the field of data-driven pedestrian trajectory prediction and simulation. The presence of Computer Science and Engineering reflects the dominance of machine learning, deep learning, and algorithmic approaches. Civil and Construction Engineering highlights the relevance of this field to infrastructure design, traffic simulation, and urban planning. Meanwhile, Mathematics \& Physics and Physical Sciences provide the theoretical and modeling foundations, such as dynamic systems and stochastic processes, essential for simulating pedestrian behavior. Environmental Science and Technology connects through applications in sustainable urban development and public space optimization.

Other disciplines, while less prominent in the visualization, also contribute to the development of pedestrian trajectory prediction and simulation. For example, Biology and Medical Sciences provide insights into human physiology and behavior, which can inform more realistic modeling of pedestrian movement, especially under stress or in emergency scenarios. Social Studies and Neurosciences offer valuable perspectives on human decision-making, social interactions, and cognitive processes, all of which are critical for simulating complex crowd dynamics. Additionally, Food and Drug Sciences appear peripherally connected, which may be linked to studies involving human movement in specific contexts such as health monitoring or public health interventions. These peripheral disciplines, though not central to the field, introduce complementary knowledge that can enrich data-driven models with more human-centered and context-aware features. Their presence on the map suggests opportunities for further interdisciplinary collaboration, particularly in addressing challenges related to public safety, health-aware mobility, and behavioral modeling within urban environments.

Overall, the Global Map of Science provides a comprehensive view of how the field of data-driven pedestrian trajectory prediction and simulation is situated within the broader scientific ecosystem, revealing both the disciplinary foundations and the opportunities for deeper interdisciplinary integration.

\subsection{R3}
To answer this question, we examine the geographical distribution, publication venues, and key contributors that collectively shape the knowledge structure of data-driven pedestrian trajectory prediction and simulation. By analyzing countries, we reveal where the main research powerhouses are located. Through the identification of influential journals and conferences, we map out the primary channels through which research outputs are disseminated. Finally, by profiling the most productive and highly cited authors, we highlight the individual and group-level contributions that have played a decisive role in setting the research agenda and advancing the state of the art.
\begin{figure}
    \centering
    \includegraphics[width=\linewidth]{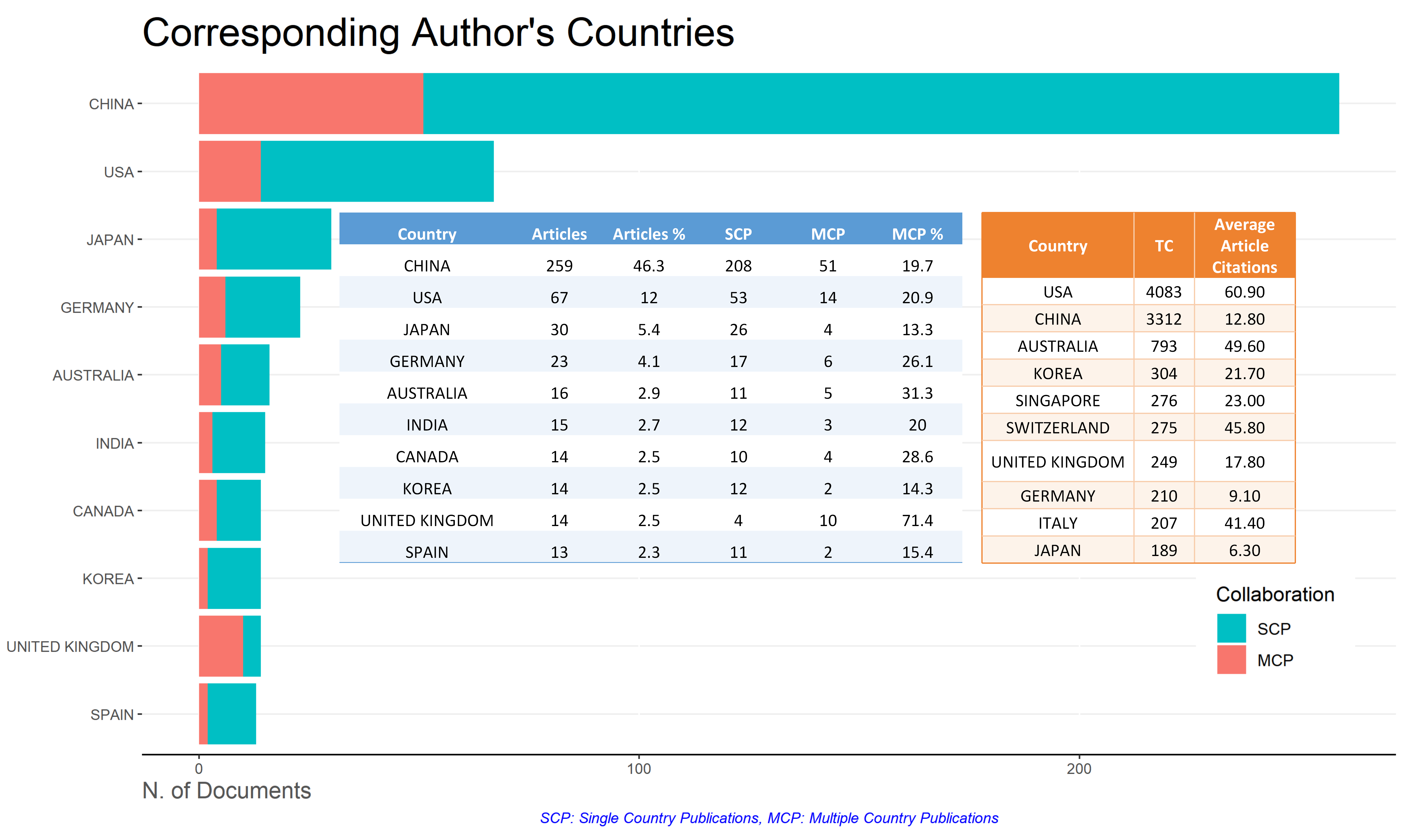}
    \caption{Geographic distribution of corresponding authors, collaboration patterns (SCP vs. MCP), and citation impact. SCP = Single Country Publications, MCP = Multiple Country Publications.}
    \label{fig:2}
\end{figure}

The Figure \ref{fig:2} illustrates the geographical distribution and collaboration patterns of corresponding authors in the dataset, highlighting both the volume of contributions and the impact in terms of total citations and average citations per article. China is the most productive country, contributing 259 articles (46.3\%), followed by the USA (67, 12\%) and Japan (30, 5.4\%). Other countries such as Germany, Australia, India, Canada, Korea, the United Kingdom, and Spain contribute smaller but notable shares. The majority of China's publications (208 out of 259) are single-country publications (SCP), with only 19.7\% involving international collaborations (MCP). In contrast, the United Kingdom exhibits a high collaboration intensity, with 71.4\% of its publications being MCP, followed by Australia (31.3\%) and Canada (28.6\%). The USA leads in both total citations (4083) and average citations per article (60.9), reflecting a combination of research quantity and influence. China, despite having the largest number of articles, has a lower average citation rate (12.8). Australia stands out with a high average of 49.6 citations per article, indicating high-quality contributions relative to its publication volume. Other countries such as Korea, Switzerland, and Singapore also show moderate to strong impact. 

\begin{table}[!ht]
\caption{\bf Distribution of articles in the dataset and the corresponding most prolific and cited sources (Top 8). N = Number of articles, TC=Total Citations.}
\footnotesize
\centering
\begin{tabular}{|p{7cm}|p{0.5cm}||p{7cm}|p{0.8cm}|p{0.4cm}|}
    \hline
    \rowcolor{orange!15}
    \multicolumn{2}{|c||}{\textbf{Most Prolific Sources
}} &
    \multicolumn{3}{c|}{\textbf{Most Cited Sources}} \\ \hline
    \rowcolor{orange!15}
    \textbf{Name} & \textbf{N} & \textbf{Name} & \textbf{TC} & \textbf{N}\\ 
     \hline
    IEEE TRANSACTIONS ON INTELLIGENT TRANSPORTATION SYSTEMS
  & 16 & 2016 IEEE/CVF CONFERENCE ON COMPUTER VISION AND PATTERN RECOGNITION (CVPR 2016)
   & 2355 & 1
  \\ \hline
    IEEE ACCESS
  & 15 & 2020 IEEE/CVF CONFERENCE ON COMPUTER VISION AND PATTERN RECOGNITION (CVPR 2020) 
   & 629 & 1
  \\ \hline
    IEEE ROBOTICS AND AUTOMATION LETTERS
   & 15 & IEEE TRANSACTIONS ON INTELLIGENT TRANSPORTATION SYSTEMS
   & 514 & 16
  \\ \hline
       COMPUTER ANIMATION AND VIRTUAL WORLDS
   & 14 & ACM TRANSACTIONS ON GRAPHICS
   & 385 & 8
  \\ \hline
       SENSORS
   & 10 & 2018 IEEE WINTER CONFERENCE ON APPLICATIONS OF COMPUTER VISION (WACV 2018)
   & 359 & 2
  \\ \hline
       EXPERT SYSTEMS WITH APPLICATIONS
   & 9 & 2019 IEEE/CVF CONFERENCE ON COMPUTER VISION AND PATTERN RECOGNITION (CVPR 2019)
      & 346 & 1
  \\ \hline
       ACM TRANSACTIONS ON GRAPHICS
   & 8 & PROCEEDINGS OF THE NATIONAL ACADEMY OF SCIENCES OF THE UNITED STATES OF AMERICA
   & 293 & 1
  \\ \hline
PHYSICA A-STATISTICAL MECHANICS AND ITS APPLICATIONS
   & 8 & NEUROCOMPUTING
   & 289 & 6
  \\ \hline
\end{tabular}
\label{tab1}
\end{table}
Table~\ref{tab1} reports the distribution of articles across publication venues, distinguishing between most prolific sources (N) and most cited sources (TC).

In terms of publication volume (N), a small group of journals and conferences dominate the field. IEEE Transactions on Intelligent Transportation Systems leads with 16 articles, followed closely by IEEE Access and IEEE Robotics and Automation Letters (15 each), and Computer Animation and Virtual Worlds (14). This reflects a clear concentration of research output in venues associated with intelligent transportation, robotics, and simulation technologies. The citation-based perspective (TC) tells a complementary story. The single most cited venue is the 2016 IEEE/CVF Conference on Computer Vision and Pattern Recognition (CVPR 2016), with 2355 citations for a single influential paper, illustrating that even a single publication in a high-impact conference can profoundly shape the field. Similar patterns are seen with CVPR 2020 and WACV 2018, which, despite having few publications, have high cumulative citation counts. These cases emphasize the disproportionate impact of top-tier computer vision conferences on the research domain. Some venues, such as IEEE Transactions on Intelligent Transportation Systems and ACM Transactions on Graphics, appear on both lists, indicating a combination of steady research output and lasting scholarly influence. Neurocomputing also illustrates sustained attention with 6 papers contributing to a TC of 289. Overall, these results reveal a dual structure in the dissemination of research:
(1) Core journals (e.g., IEEE T-ITS, IEEE Access) provide a consistent publication platform and accumulate influence through steady contributions;
(2) High-impact conferences (e.g., CVPR) serve as catalysts for breakthrough innovations, where a few landmark papers garner exceptional citations.
\begin{table}[!ht]
\caption{\bf Author statistics in the dataset, including productivity (N), research influence (H-index), and total citations (TC). 
N = Number of articles; TC = Total Citations.}
\footnotesize
\centering
\begin{tabular}{|p{3.5cm}|p{0.8cm}||p{3.5cm}|p{1cm}||p{3.5cm}|p{1cm}|p{0.8cm}|}
    \hline
    \rowcolor{red!15}
    \multicolumn{2}{|c||}{\textbf{Most Prolific Authors}} &
    \multicolumn{2}{c||}{\textbf{Top H-Index Authors}} &
    \multicolumn{3}{c|}{\textbf{Most Cited Authors}} \\ \hline
    \rowcolor{red!15}
    \textbf{Author} & \textbf{N} &
    \textbf{Author} & \textbf{H-Index} &
    \textbf{Author} & \textbf{TC} & \textbf{N}\\ 
    \hline
    KAPADIA, MUBBASIR & 9 & LIU HONG & 6 & ALAHI ALEXANDRE & 2574 & 4 \\ \hline
    MANOCHA, DINESH & 7 & MANOCHA DINESH & 6 & GOEL KRATARTH & 2355 & 1 \\ \hline
    LIU, HONG & 6 & ZHENG NANNING & 6 & LI FEI-FEI & 2355 & 1 \\ \hline
    MAO, TIANLU & 6 & CAI WENTONG & 5 & RAMANATHAN VIGNESH & 2355 & 1 \\ \hline
    PAVLOVIC, VLADIMIR & 6 & ALAHI ALEXANDRE & 4 & ROBICQUET ALEXANDRE & 2355 & 1 \\ \hline
    WANG, ZHAOQI & 6 & BERA ANIKET & 4 & SAVARESE SILVIO & 2355 & 1 \\ \hline
    ZHENG, NANNING & 6 & CHEN WANGXING & 4 & QIAN KUN & 642 & 2 \\ \hline
    BERA, ANIKET & 5 & FALOUTSOS PETROS & 4 & CLAUDEL CHRISTIAN & 629 & 1 \\ \hline
    CAI, WENTONG & 5 & KAPADIA MUBBASIR & 4 & ELHOSEINY MOHAMED & 629 & 1 \\ \hline
    CHEN, MO & 5 & SANG HAIFENG & 4 & MOHAMED ABDUALLAH & 629 & 1 \\ \hline
\end{tabular}
\label{tab:authors}
\end{table}

Table~\ref{tab:authors} presents a comprehensive analysis of authorship patterns in the dataset across three complementary dimensions: productivity (N), academic influence (H-index), and total citations (TC).

From the perspective of productivity, the field shows a concentrated distribution with a small group of highly active authors. Kapadia, Mubbashir leads with 9 publications, followed by Manocha, Dinesh with 7, while several other scholars (e.g., Liu Hong, Mao Tianlu, Pavlovic Vladimir, Wang Zhaoqi, Zheng Nanning) each contributed 6 articles. Academic influence, as measured by the H-index, highlights authors whose contributions are not only frequent but also sustained in terms of citation relevance. Liu Hong, Manocha Dinesh, and Zheng Nanning achieve the highest H-index (6), indicating consistent recognition of their works within the community. In contrast, the total citations (TC) column captures cumulative impact rather than consistency. Alahi Alexandre, for instance, has accumulated 2574 citations across just four articles, showing that a small number of high-impact papers can disproportionately influence the field. Similarly, authors such as Goel Kratarth, Li Fei-Fei, and Ramanathan Vignesh are associated with single highly cited papers. These high TC values should therefore be interpreted as indicators of landmark contributions, not overall productivity.

These results collectively reveal a three-tiered structure of scholarly contribution in the field: (1) a core group of prolific authors driving research volume, (2) a partially overlapping group of authors with strong H-index values representing sustained academic influence, and (3) a set of researchers with fewer but highly impactful contributions that attract large citation counts. This structure suggests that while the field benefits from a steady output of contributions, its intellectual advancement is also shaped by a few landmark works.

\subsection{R4}
\begin{figure}
    \centering
    \includegraphics[width=\linewidth]{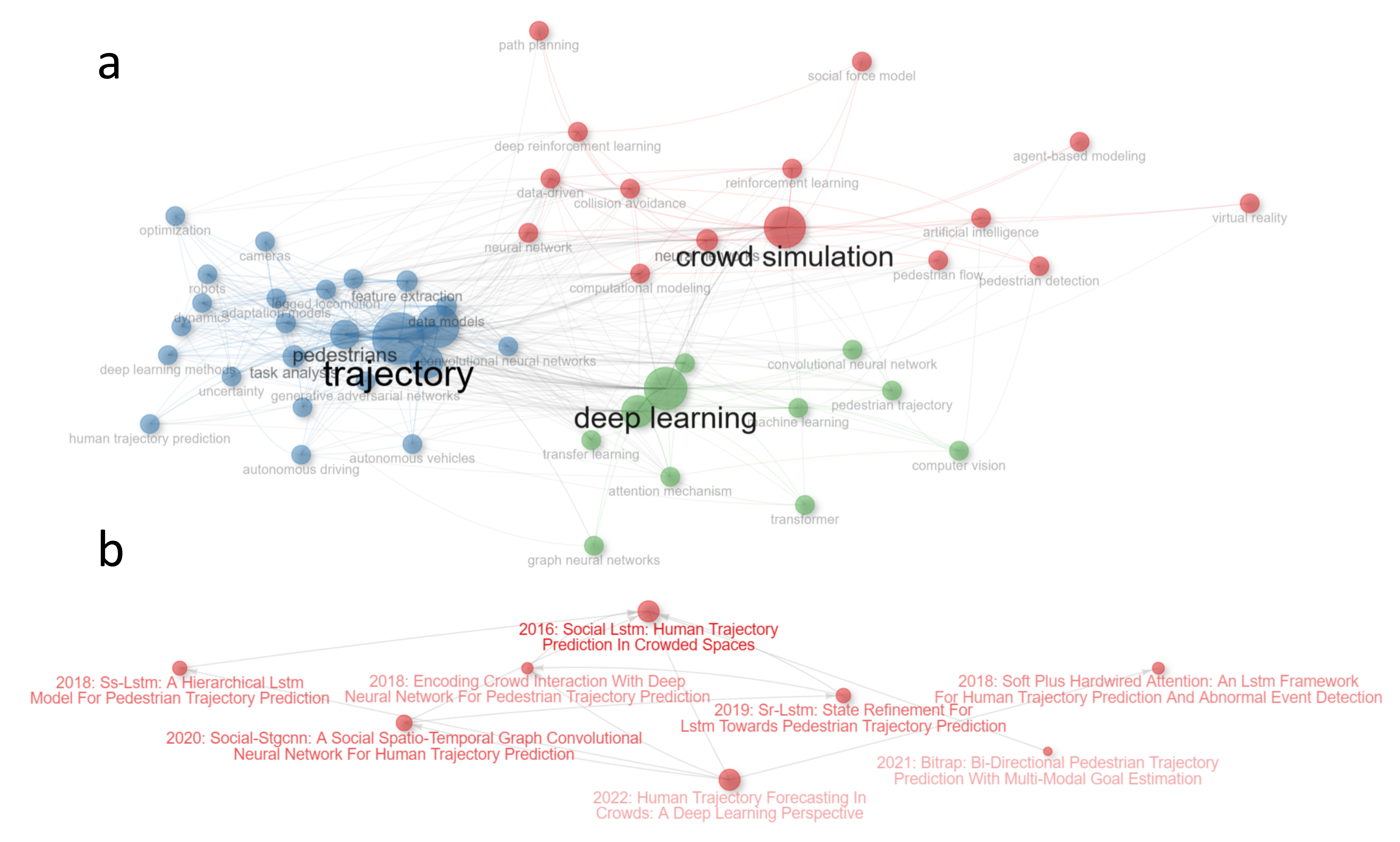}
    \caption{Research Hotspots and Intellectual Evolution in Data-Driven Pedestrian Trajectory Prediction and Simulation.}
    \label{fig:4}
\end{figure}
As shown in Figure \ref{fig:4}, the keyword co-occurrence network (a) reveals three tightly connected thematic clusters representing the core research directions within the field. The blue cluster centers on "trajectory" and "pedestrians", and is densely connected to keywords such as "autonomous vehicles", "generative adversarial networks", and "human trajectory prediction", reflecting a strong emphasis on machine learning-based trajectory modeling with applications in autonomous driving and robotics. The green cluster is dominated by "deep learning", linking to contemporary techniques like transformers, graph neural networks, and attention mechanisms, indicating an increasing methodological shift from traditional models to data-driven architectures. In contrast, the red cluster led by "crowd simulation" is strongly associated with reinforcement learning, path planning, agent-based modeling, and collision avoidance. Notably, it includes classical frameworks such as the Social Force Model (SFM)—a physics-inspired approach where pedestrian movement is modeled as particles subject to social and physical forces (e.g., repulsion from obstacles or attraction to goals). While SFM has been foundational for simulating collective pedestrian behavior in crowded environments, its integration with deep learning remains limited, suggesting a potential research gap in unifying realistic physical modeling and data-driven learning in multi-agent crowd scenarios.
\begin{table}[!ht]
\caption{\bf Top 5 cited works in the dataset, including title, source, document type, year of publication, total citations (TC), and authors.}
\footnotesize
\centering
\begin{tabular}{|p{4cm}|p{2cm}|p{1.2cm}|p{1.2cm}|p{1.2cm}|p{5cm}|}
    \hline
    \rowcolor{green!15}
    \textbf{Title} & \textbf{Source} & \textbf{Type} & \textbf{Year} & \textbf{TC} & \textbf{Authors} \\ \hline
    
    Social LSTM: Human trajectory prediction in crowded spaces \cite{alahi2016social}& 
    2016 IEEE/CVF CVPR & 
    Proc. Paper & 
    2016 & 
    2355 & 
    Alahi, Alexandre; Goel, Kratarth; Ramanathan, Vignesh; Robicquet, Alexandre; Li, Fei-Fei; Savarese, Silvio \\ \hline
    
    Social-STGCNN: A social spatio-temporal graph convolutional neural network for human trajectory prediction \cite{mohamed2020social}& 
    2020 IEEE/CVF CVPR & 
    Proc. Paper & 
    2020 & 
    629 & 
    Mohamed, Abduallah; Qian, Kun; Elhoseiny, Mohamed; Claudel, Christian \\ \hline
    
    SR-LSTM: State refinement for LSTM towards pedestrian trajectory prediction \cite{zhang2019sr}& 
    2019 IEEE/CVF CVPR & 
    Proc. Paper & 
    2019 & 
    346 & 
    Zhang, Pu; Ouyang, Wanli; Zhang, Pengfei; Xue, Jianru; Zheng, Nanning \\ \hline
    
    SS-LSTM: A hierarchical LSTM model for pedestrian trajectory prediction \cite{xue2018ss}& 
    2018 IEEE WACV & 
    Proc. Paper & 
    2018 & 
    307 & 
    Xue, Hao; Huynh, Du Q.; Reynolds, Mark \\ \hline
    
    Social behavior for autonomous vehicles \cite{schwarting2019social}& 
    PNAS & 
    Article & 
    2019 & 
    293 & 
    Schwarting, Wilko; Pierson, Alyssa; Alonso-Mora, Javier; Karaman, Sertac; Rus, Daniela \\ \hline
\end{tabular}
\label{tab:top5cited}
\end{table}

Building on these thematic insights, we further examined the intellectual foundations of the field by identifying the most influential works associated with these clusters. Specifically, after confirming that the convergence of machine/deep learning and pedestrian-related research constitutes the central axis of recent developments, we extracted the five most highly cited papers from the dataset. The top five most cited works (see Table \ref{tab:top5cited}) reflect the methodological evolution and application breadth of pedestrian trajectory prediction research. The earliest milestone, Social-LSTM (2016), marks the transition from handcrafted rules to deep learning-based approaches, laying the foundation for subsequent innovations. Works such as SS-LSTM (2018), SR-LSTM (2019), and Social-STGCNN (2020) represent progressive refinements in recurrent, hierarchical, and graph-based modeling of spatio-temporal interactions, emphasizing increasingly structured and multimodal representations. In parallel, the inclusion of the PNAS article Social behavior for autonomous vehicles (2019) demonstrates the field’s expansion toward safety-critical applications, bridging human crowd dynamics with autonomous navigation. Collectively, these papers illustrate not only the dominance of LSTM-derived frameworks in early deep learning trajectories but also a growing shift toward graph and socially-aware models with strong practical relevance. Notably, most of these highly cited works were published in top-tier computer vision and artificial intelligence conferences rather than journals. This concentration in conference proceedings reflects the rapid pace of methodological turnover in the field: the long review and publication cycles of traditional journals struggle to match the accelerated tempo of innovation, prompting researchers to prioritize conference venues that ensure faster dissemination and greater visibility. The dominance of flagship conferences such as CVPR and WACV thus underscores the dynamic, fast-evolving nature of pedestrian trajectory prediction research at the intersection of deep learning and autonomous systems.

Using the most cited of these as a reference point, we then traced its citation trajectory across subsequent studies to reconstruct the chronological research lineage. This approach allows us to contextualize the evolution of methodological innovations and application foci, while highlighting how a single seminal contribution has catalyzed diverse follow-up studies and shaped the broader research landscape. Among them, Social-LSTM (2016) stands out as the most highly cited work, authored by Alexandre Alahi, Kratarth Goel, Vignesh Ramanathan, Alexandre Robicquet, Fei-Fei Li, and Silvio Savarese. Social-LSTM introduced a pioneering recurrent neural network framework that explicitly modeled human trajectories in crowded spaces by incorporating social pooling, enabling the network to capture interdependent motion patterns among pedestrians. This methodological breakthrough not only shifted trajectory prediction away from handcrafted rules toward data-driven deep learning approaches, but also established a foundation upon which nearly all subsequent research has been built. The prominence of its authors and its transformative impact explain why Social-LSTM continues to serve as a cornerstone in the literature, frequently cited as the canonical starting point for modern trajectory prediction research. Notably, Fei-Fei Li and Silvio Savarese are widely recognized figures in AI, whose prior contributions to large-scale visual recognition and scene understanding have profoundly shaped the field. Recently, Li has spearheaded the development of “spatial intelligence” through her startup World Labs, aiming to equip AI with a richer understanding of 3D environments—a direction that promises to influence future trajectory and crowd modeling methods.

Anchored at Social-LSTM (2016), the lineage in Figure \ref{fig:4} (b) shows a clear methodological arc from recurrent modeling of social interactions to structured graphs and multimodal goal-conditioning. Social-LSTM introduced the now-canonical social pooling to jointly reason over neighboring agents, replacing hand-crafted interaction rules and establishing a data-driven baseline for crowded scenes. In 2017/2018, two LSTM branches deepened interaction modeling: Soft+Hardwired Attention  fused learned (“soft”) and geometry-aware (“hard-wired”) attention to weight neighbors at scale, while SS-LSTM made the architecture hierarchical to inject scene/layout context alongside social cues—both addressing limitations of uniform pooling and scene-agnostic models. A parallel 2018 thread, Encoding Crowd Interaction with Deep Neural Network (CIDNN), proposed spatial-affinity weighting and displacement (residual) prediction between frames, arguing that even distant fast-moving agents can influence the target pedestrian; this shifted attention from fixed-neighborhood heuristics to learned pairwise influence.

Refinement and structure then dominate. SR-LSTM (2019) introduces state-refinement message passing among LSTM nodes to iteratively polish hidden states, improving interaction fidelity under dense crowds. Social-STGCNN (2020) recasts trajectories as a spatio-temporal graph and learns interactions with graph convolutions, reporting large FDE/ADE gains with far fewer parameters and faster inference—evidence that structured relational inductive biases outperform ad-hoc aggregation. The focus then pivots to multimodality and intent: BiTraP (2021) couples a conditional variational autoencoder (CVAE) with goal-conditioned bi-directional decoding to better handle long-horizon uncertainty and endpoint intent across  first-person view (BEV) and bird’s-eye view (FPV) settings. 

Finally, the 2022 work Human Trajectory Forecasting in Crowds: A Deep Learning Perspective contributes two knowledge-based data-driven modules, the TrajNet++ interaction-centric benchmark, and socially-aware metrics, pushing the field toward standardized, interaction-sensitive evaluation rather than raw displacement error alone. 

Viewed together, the succession is coherent: (i) learn social context (Social-LSTM) → (ii) weight it better and add scene priors (Soft+Hardwired, SS-LSTM, CIDNN) → (iii) refine interactions via message passing (SR-LSTM) → (iv) structure them as graphs for efficiency and generality (Social-STGCNN) → (v) model multimodal intent (BiTraP) → (vi) institutionalize evaluation with interaction-centric benchmarks and metrics (TrajNet++). This trajectory also clarifies near-term gaps: bridging graph/intent models with simulation-level controls (e.g., reinforcement-learning planners) and adopting social-metric-aligned objectives during training, so that learned predictors not only minimize error but also conform to human-acceptable behaviors in dense, multi-agent scenes.

\section{R5}
Since our dataset extends only through 2025, it is unsurprising that works published in 2024 and 2025 may not yet have accrued sufficient citations to emerge among the top-cited studies. Nevertheless, these newer contributions often embody cutting-edge methodologies and innovative modalities that are poised to become the field’s next milestones. 
\begin{table}[!ht]
\caption{\bf The 5 most cited works for each of the years 2024 and 2025 in the dataset, including title, source, year of publication, total citations (TC), and authors.}
\footnotesize
\centering
\begin{tabular}{|p{4.5cm}|p{4cm}|p{1cm}|p{0.5cm}|p{5cm}|}
    \hline
    \rowcolor{blue!15}
    \textbf{Title} & \textbf{Source} & \textbf{Year} & \textbf{TC} & \textbf{Authors} \\ \hline
    
    Meta-IRLSTP++: A meta-inverse reinforcement learning method for fast adaptation of trajectory prediction networks \cite{yang2024meta}& 
    Expert Systems with Applications & 

    2024 & 
    23 & 
    Yang, Biao; Lu, Yanan; Wan, Rui; Hu, Hongyu; Yang, Changchun; Ni, Rongrong \\ \hline
    
    STGlow: A flow-based generative framework with dual-graphormer for pedestrian trajectory prediction \cite{liang2024stglow}& 
    IEEE Trans. on Neural Networks and Learning Systems & 

    2024 & 
    16 & 
    Liang, Rongqin; Li, Yuanman; Zhou, Jiantao; Li, Xia \\ \hline
    
    Spatial-Temporal-Spectral LSTM: A transferable model for pedestrian trajectory prediction \cite{zhang2024spatial}& 
    IEEE Trans. on Intelligent Vehicles & 
 
    2024 & 
    12 & 
    Zhang, Chi; Ni, Zhongjun; Berger, Christian \\ \hline
    
    IA-LSTM: Interaction-aware LSTM for pedestrian trajectory prediction \cite{yang2024ia}& 
    IEEE Trans. on Cybernetics & 
   
    2024 & 
    12 & 
    Yang, Jing; Chen, Yuehai; Du, Shaoyi; Chen, Badong; Principe, Jose C. \\ \hline
    
    STIGCN: Spatial-temporal interaction-aware graph convolution network for pedestrian trajectory prediction \cite{chen2024stigcn}& 
    Journal of Supercomputing & 
    
    2024 & 
    10 & 
    Chen, Wangxing; Sang, Haifeng; Wang, Jinyu; Zhao, Zishan \\ \hline
    
    Disentangling the hourly dynamics of mixed urban function: A multimodal fusion perspective using dynamic graphs \cite{cao2025disentangling}& 
    Information Fusion & 
    
    2025 & 
    5 & 
    Cao, Jinzhou; Wang, Xiangxu; Chen, Guanzhou; Tu, Wei; Shen, Xiaole; Zhao, Tianhong; Chen, Jiashi; Li, Qingquan \\ \hline
    
    Pedestrian trajectory prediction using goal-driven and dynamics-based deep learning framework \cite{wang2025pedestrian}& 
    Expert Systems with Applications & 
    
    2025 & 
    3 & 
    Wang, Honghui; Zhi, Weiming; Batista, Gustavo; Chandra, Rohitash \\ \hline
    
    Generating natural pedestrian crowds by learning real crowd trajectories through a transformer-based GAN \cite{yan2025generating}& 
    Visual Computer & 
   
    2025 & 
    3 & 
    Yan, Dapeng; Ding, Gangyi; Huang, Kexiang; Huang, Tianyu \\ \hline
    
    Coordinating dynamic signage for evacuation guidance: A multi-agent reinforcement learning approach integrating mesoscopic crowd modeling and fire propagation \cite{xie2025coordinating}& 
    Chaos, Solitons \& Fractals & 
    
    2025 & 
    2 & 
    Xie, Chuan-Zhi Thomas; Chen, Qihua; Zhu, Bin; Lee, Eric Wai Ming; Tang, Tie-Qiao; Yin, Xianfei; Yuan, Zhilu; Zhang, Botao \\ \hline
    
    Efficient crowd simulation in complex environment using deep reinforcement learning \cite{li2025efficient}& 
    Scientific Reports & 
    
    2025 & 
    2 & 
    Li, Yihao; Chen, Yuting; Liu, Junyu; Huang, Tianyu \\ \hline
\end{tabular}
\label{tab:recent_topcited}
\end{table}

As shown in Table \ref{tab:recent_topcited}, the most cited works of 2024–2025 are not merely extensions of existing models but herald novel methodological shifts with significant implications for the future of trajectory prediction and simulation. The 2024 set is dominated by works focused on fast adaptation, generative modeling, and generalization, reflecting the field's maturation: Meta-IRLSTP++ embeds meta-learning within an inverse reinforcement learning framework, enabling rapid adaptation to new scenes with limited data—addressing a key challenge in real-world deployment. STGlow introduces a flow-based generative model with dual graphormer architecture, optimizing for exact log-likelihood and delivering physically interpretable prediction dynamics. The trio of Spatial-Temporal-Spectral LSTM, IA-LSTM, and STIGCN focus on transfer learning, interaction-aware recurrent structures, and graph convolutional modeling, respectively—strengthening robustness and multi-agent reasoning in trajectory forecasting. For 2025, although citations are lower given recency, the selected works point toward several emergent frontiers: \citeauthor{cao2025disentangling} apply dynamic graph-based multimodal fusion to capture hourly urban dynamics, suggesting a move toward spatiotemporal semantic context. \citeauthor{yan2025generating} utilize a transformer-based GAN to generate realistic crowd behavior, signaling deeper integration of generative modeling with trajectory data. Others explore goal-driven dynamics, reinforcement learning for control-oriented simulation, and RL-based crowd simulation in complex environments, pushing toward domains where prediction meets planning and control. 

Beyond algorithmic improvements, future pedestrian intelligence research must account for social diversity, cultural behavior, and ethical use of mobility data. Cities differ not only in geometry but in social norms; integrating such contextual information is essential for globally transferable and equitable models.

\section{Discussion and Conclusion}
This study provides a quantitative and conceptual mapping of data-driven pedestrian trajectory prediction and simulation, offering insights into its evolution from physics-based models to deep learning, graph reasoning, and generative paradigms. While the scientometric evidence underscores rapid methodological progress, the broader implications for urban systems are equally profound.

\textbf{Urban and Societal Implications.} The rise of AI-driven pedestrian modeling has begun to reshape how cities perceive and manage mobility. Predictive models are no longer confined to academic laboratories; they now inform digital twins, evacuation systems, and pedestrian-oriented design policies. As urban environments become more complex, integrating these models into real-world planning supports adaptive crowd management, improves infrastructure efficiency, and enhances safety during large-scale events or emergencies. Yet, their deployment must be guided by transparency and inclusivity to prevent algorithmic bias from reinforcing spatial or social inequalities in public spaces.

\textbf{From Data-Driven to Human-Centered Urban Intelligence. }The scientometric findings reveal a disciplinary shift—from isolated computer vision studies to integrative frameworks that fuse human behavior modeling, sensor data, and urban analytics. This evolution signals the emergence of a new research paradigm: urban mobility intelligence. Future work should bridge prediction and control by embedding human factors, ethics, and behavioral diversity into algorithmic models, ensuring that technical progress translates into socially responsible urban outcomes.

\textbf{Ethical and Equity Considerations.} While AI-powered pedestrian prediction holds potential for safer and more efficient cities, it also raises ethical questions regarding data privacy, representational bias, and inclusivity. Datasets often reflect limited socio-cultural contexts, potentially reinforcing inequalities when applied globally. Future frameworks should thus incorporate fairness-aware modeling and participatory data governance to ensure that predictive systems support, rather than marginalize, diverse urban populations.

\textbf{Research Frontiers and Future Directions.} The most recent works (2024–2025) highlight three converging directions:
(1) Adaptivity and Generalization through meta-learning and transfer frameworks for deployment across diverse urban contexts;
(2) Interpretability and Trustworthiness, enabling policymakers and citizens to understand and trust AI-driven mobility systems; and
(3) Simulation–Decision Integration, where predictive models guide real-time urban control and emergency planning.
These trends collectively point toward a future in which pedestrian intelligence becomes a cornerstone of smart, equitable, and resilient cities.

\textbf{Policy and Governance Relevance.} The insights derived from this scientometric mapping can guide policymakers and urban planners in prioritizing data infrastructures and interdisciplinary collaborations. By identifying where research momentum is concentrated—such as graph-based mobility modeling or human-centered crowd simulation—cities can allocate resources more effectively toward predictive safety systems, sustainable mobility infrastructure, and real-time digital governance. Bridging research and policy in this manner is vital for translating technological capability into socially beneficial urban outcomes.

\textbf{Conclusion.} By situating data-driven pedestrian trajectory research within the broader discourse of urban intelligence, this work demonstrates that advances in AI and modeling can meaningfully inform the design of human-centered, adaptive cities. The integration of technical innovation with social responsibility will be central to realizing the promise of truly intelligent and inclusive urban mobility systems.

\section{Data availability}
All bibliometric data are available from the Web of Science Core Collection under institutional access, and the processed dataset used in this study can be obtained upon request from the corresponding author.

\bibliographystyle{elsarticle-harv} 
\bibliography{reference}

\end{document}